\documentclass[a4paper,12pt]{article}

\begin{document}
\section*{\uppercase{On The existence Of A Non-Relativistic Hermitian Time Operator}}

Kinyanjui Carringtone,Simiyu Dismas Wamalwa\\
\textit{School of Physical Sciences,University of Nairobi} \\ \\
 \ \ The existence of a hermitian time operator is proposed in the framework of non-relativistic   quantum mechanics.The Heisenberg equation of motion is shown to yield constant rate of flow of time.It is shown to yield results consistent with classical expectations,when Bohr's correspondence principle is invoked.Further,it is shown that a massless free particle with non-zero momentum experiences timelessness in its own frame  of reference whereas a negatively massive particle perceives forwardness of time.Moreover,application to quantum harmonic oscillators shows instantaneous transition of electrons from one eigenstate to another.\\ \\

\textbf{Introduction}\\ \\
In standard quantum mechanics measurable properties of a system are usually assigned with  Hermitian operators$^1$.Usually time is not taken to be an internal property of quantum systems because no such operator,hermitian in nature, has been found to be viable in the treatment of quantum systems$^{2}$.We seek to develop such an operator,without altering the basic mathematical structure and physical postulates of quantum theory.We hold that time can be,and in fact is a property of systems in quantum mechanics.
\section{Derivation}
Assuming (due to the correspondence principle),$^5$ that operators have relations analogous to the classical properties they represent,we take account of the kinetic energy operator -: $^1$
\begin{eqnarray}
\hat{T} = -\frac{\hbar^2}{2m}\frac{\partial^2}{\partial x^2}
\end{eqnarray}
We can equate this to the classical expression of kinetic energy to obtain a velocity operator -:

\[\frac{1}{2}m\hat{v}^2 = -\frac{\hbar^2}{2m}\frac{\partial^2}{\partial x^2}\]

\[\hat{v}^2 = -\frac{\hbar^2}{m^2}\frac{\partial^2}{\partial x^2}\]
\begin{eqnarray}
\hat{v} = \pm \frac{i\hbar}{m}\frac{\partial^2\psi}{\partial x^2}
\end{eqnarray}
From classical dynamics we know that velocity and displacement are considered to be vectors and time is a scalar.Since we expect the operator to reproduce the results of classical dynamics in large quantum numbers and we also expect that the time operator be hermitian in nature,we can select the negative components of the velocity operator.

\subsection{The displacement operator}
We postulate an operator $\hat{D}$ of the form -:
\begin{eqnarray}
\hat{D} = -\frac{\hbar^2}{p^2}\frac{\partial^2}{\partial x^2}
\end{eqnarray}

The justification we give for this is that it yields the classical result for distances when we consider many body systems.Therefore its expectation value in a state $\psi$ is given by\\ \\

$<D> = \int \psi \hat{D}\psi dx$ \\ \\
If we consider a free particle system such that \ \

$\displaystyle \psi=A e^{\frac{i(px-Et)}{\hbar}}$\\ \\ 

Setting upper and lower bounds at $b$ and $a$ respectively -:
\[<D> = \int _a^b e^{\frac{i(Et-px)}{\hbar}}e^{\frac{i(px-Et)}{\hbar}} dx\]
\begin{eqnarray}
<D> = b-a 
\end{eqnarray}

Since $x$ is in units of distance, $b-a$ is in units of distance consistent with the classical description of the same Bohr's correspondence principle.Now for a particle of mass $m$ and momentum eigenvalues $p$ we can use equations (2) and (3), to write down the one dimensional hermitian time operator as \\ \\
\begin{eqnarray}
\displaystyle \hat{t}=\frac{\hbar m}{ip^2}\frac{\partial}{\partial x} 
\end{eqnarray}

\section{Consistency tests}
We now seek to test the operator in equation (5) for consistency with standard quantum theory.We check it for hermicity,consistency with Bohr's correspondence principle and Heisenberg's equation of motion.
\subsection{Proof of hermicity}
If $t$ is hermitian,then

\[<\hat{t}f |g> = <f|\hat{t}g>\]
Or
\[<\hat{t}f |g> = \int_a^b\left(\frac{i\hbar m}{p^2}\frac{\partial f}{\partial x}\right)^\ast gdx\]
Assuming f and g vanish at a and b respectively-:
\[<\hat{t}f|g>=\int_a^b\frac{ \partial f^*}{\partial x} \left(\frac{-i\hbar m}{p^2}g\right)dx \]
Integrating the above equation by parts and assuming f and g are analytic
functions of x so that for large x, a and b tend towards minus and plus infinity
respectively so that f, g vanishes we have
\[<\hat{t}f|g> =\int_a^b f\left(\frac{i\hbar m}{p^2}\frac{\partial g}{\partial x}\right)^*dx\]
\begin{eqnarray}
<\hat{t}f|g> = <f|\hat{t}g>
\end{eqnarray}
 therefore $ \hat{t}$ is hermitian

\subsection{Test of consistency with Bohr's correspondence principle}
In large quantum numbers we expect quantum theory to reproduce results of classical mechanics according to Bohr's correspondence principle $^{5}$.
The expectation value for time for a free particle is given as -:
\begin{eqnarray}
<t> & = & \int_a^b \psi^\ast\frac{-i\hbar m}{p^2}\frac{\partial}{\partial x}\psi dx \nonumber \\
<t> & = & \frac{i\hbar m}{p^2} \int _a^b \psi^\ast \frac{\partial}{\partial x}\psi dx \nonumber \\
<t> & = & \frac{m}{p}\int_a^b e^{\frac{i(Et-px)}{\hbar}} e^{\frac{i(px-Et)}{\hbar}} dx \nonumber \\
<t> &=& \frac{m\left(b-a \right)}{p}\nonumber \\
<t> &=& \frac{b-a}{v} = t 
\end{eqnarray}
In large quantum numbers the value of $p$ tends towards $mv$ 

\subsection{Application of heisenberg's equation of motion to the hermitian time operator}
Heisenberg's equation of motion is $^{6}$-:
\begin{eqnarray}
\frac{d<A>}{dt} = i\hbar \left[\hat{A}, \hat{H}\right] + \frac{\partial \hat{A}}{\partial t} 
\end{eqnarray}

Where $\hat{A}$ is a general operator.
We seek to use equation(8) to help us understand the flow of time in non-relativistic quantum systems.By inspection we expect that

$\displaystyle \frac{d<t>}{dt} = 0$ 

i.e the expected value for time in a physical process does not change.Equation (8) can be written in terms of the time operator as:-\\ \\
\begin{eqnarray}
 \frac{d}{dt}<t> = i\hbar \left[\hat{t}, \hat{H}\right] + \frac{\partial \hat{t}}{\partial t}
\end{eqnarray}
 
If we consider a one dimensional linear harmonic oscillator at zero potential whose state is time independent it is easy to show from equations (5) and (9) that
\begin{eqnarray}
\frac{d<t>}{dt} = 0
\end{eqnarray}

The rate of flow of time is constant in quantum mechanics for the non-relativistic case.This is the quantum mechanical way of stating that time is absolute the in non-relativistic case$^7$.

\section{Application to idealised physical systems}
\subsection{The free particle system}
According to Newton's law of motion$^3$ for a particle under no external forces will take an infinite amount of time to stop or change direction.We seek to explore the same situation quantum-mechanically by use of the time operator and standard Sturm-Liouville methods.In Sturm-Liouville theory $^8$ the expectation value of an observable E is represented by -:\\ $\displaystyle<E> = \int_{-\infty}^{+\infty}\psi^\ast\hat{E}\psi dv $ \\
Applying this to the time operator treatment of the free particle in the one dimensional case leads to-:
\begin{eqnarray}
<t> &=& \int_{-\infty}^{+\infty} \psi^\ast \hat{t}\psi dv\nonumber \\
<t> &=& \infty 
\end{eqnarray}

Therefore the expectation value for time for a free particle is infinity consistent with classical mechanics.

\subsection{A massless free particle}
Consider a massless beam of particles created at $a$ and absorbed at $b$.We can calculate the expectation value for time taken by the particle between emission and absorption using equations (4) and (5) as-:
\begin{eqnarray}
<t> &=& \int \psi^\ast \cdot\hat{t}\psi dx\ \nonumber \\
&=&\frac{m}{p} \left[x\right]_ a ^b \nonumber
\end{eqnarray}
For $m=0$ and $p\neq0$;
\begin{eqnarray}
<t>=0
\end{eqnarray}

Though we sought to consider the problem non-relativistically,we can see that this is consistent with Einstein's special theory of relativity$^3$ where -: 
$\displaystyle \bigtriangleup t^{'} =\gamma \bigtriangleup t$
Where $\gamma$ is the Lorentz factor given by $\displaystyle \frac{1}{\sqrt{1-\frac{v^2}{c^2}}•}
$ so that\\$\displaystyle \bigtriangleup t\rightarrow 0 $ when $v \rightarrow c$ as applies to the photon.
\subsection{Consideration of a free particle \\with negative rest mass}
If we now consider a particle with negative rest mass we may anticipate the operator to yield backward flow of time.However, considering the problem as before for a free particle it is easy to show using equations(4) and (5) that-:
\begin{eqnarray}
<\hat{t}> &=& \int_{-\infty}^{+\infty} \psi^\ast\hat{t}\psi dx\nonumber \\
 <\hat{t}>&=&\infty
\end{eqnarray}
since m is negative by definition.The value is still positive.This only points to the fact that time can only flow forwards for systems locally.Backward flow of time seems to be a local impossibility.

\section{Application of the operator to a \newline real physical system}
\subsection{The Quantum Harmonic Oscillator}
In calculation of expectation values in the situation of quantum harmonic oscillators we  exploit the use of ladder operators $^1$ to simplify the situation under consideration.

\subsubsection{Expressing the time operator in terms of ladder operators}
We make use of the relationship between the time operator and the momentum operator to construct a hermitian time operator expressible in terms of ladder operators.Given that-:

\[\hat{p} \rightarrow-i\hbar \frac{\partial \psi}{\partial x}  \ \ \ \ ^8\] 
\[\hat{t} \rightarrow\frac{\hbar m}{ip^2} \frac{\partial \psi}{\partial x}\]

And that for a linear harmonic oscillator the momentum operator is written in terms of ladder operators as-:

\begin{eqnarray}
\hat{p} = i\sqrt{\frac{m\hbar\omega}{2}}\left(a -a^\dag \right) \ \ \ ^8
\end{eqnarray}

Where $a$ is a lowering operator and $a^\dag$ is a raising operator.We exploit the relationship between the time operator and the operator in equation(14) to find an expression for time operator in terms of ladder operators.

If\ $\displaystyle\hat{p}\psi = \frac{p^2}{m} \hat{t}\psi$ then-:
\begin{eqnarray}
\hat{t}\psi &=& \frac{m}{p^2} \hat{p} \psi\nonumber\\
\Rightarrow\hat{t}\psi &=& \frac{im}{p^2}\sqrt{\frac{m\hbar\omega}{2}} \left(a - a^\dag \right) \psi 
\end{eqnarray}
We can then use this operator in the standard way to find the expectation value for time in a quantum harmonic oscillator. 

\subsubsection{\textit{Finding the expectation value for time in a quantum\newline harmonic oscillator when acted on by ladder operators}}
The expectation value for time in a quantum harmonic oscillator is
\begin{eqnarray}
<\hat{t}> &=& <n |\hat{t}| n>\nonumber \\
<\hat{t}> &=& <n |\frac{im}{p^2} \sqrt{\frac{m\hbar\omega}{2}} \left(a - a^+ \right) | n>\nonumber \\
<\hat{t}> &=& \frac{im}{p^2}\sqrt{\frac{m\hbar\omega}{2}} \left(<n | a^+ |n> - <n |a| n> \right)\nonumber
\end{eqnarray}
Let  \ \ $\displaystyle \frac{im}{p^2} \sqrt{\frac{m\hbar\omega}{2}}= \alpha$
\begin{eqnarray}
<\hat{t}> &=& \alpha \left(<n |\sqrt{n+1}| n+1> - <n |\sqrt{n}| n-1> \right)\nonumber \\
<\hat{t}> &=& \alpha \left(<n |n+1> \sqrt{n+1} - \sqrt{n} <n |n-1> \right)\nonumber
\end{eqnarray}
Since state n is orthonormal to state $n+1$ and state $n-1$
\begin{eqnarray}
<n |n+1> &=& <n |n-1> = 0\nonumber \\
<t> &=& \alpha \cdot 0\nonumber \\
<t> &=& 0
\end{eqnarray}
\subsubsection{Interpretation}
We propose that an interpretation of equation(16) can be found in the description of quantum jumps.This result lends mathematical support to the assumption by Bohr that transitions between states are instantaneous.$^{9}$ We shall use this fact to make a principally testable hypothesis.

\section{Prediction}
Since the operator predicts instantaneity $^9$ in electron transition between states, no time of transition between energy states is expected.Any such times of transitions must be attributed to perturbations in the systems,chief of which is Heisenberg's inequality for energy $^8$-: 
\begin{eqnarray}
\Delta E \cdot \Delta t \geq \frac{\hbar}{2}
\end{eqnarray}

Where $\Delta E$ is the energy of the particle after action on it by a ladder operator. We can use (17) to construct a further inequality-:
\begin{eqnarray}
\tau_J \leq \Delta t
\end{eqnarray}

The $ jump\ \ time$ as defined by Schulman L,$^{2} (\tau_J)$ must be less than or equal to time
$\Delta t$ as a result of Heisenberg's inequality.If inequality $(18)$ is violated, then the operator does not hold true.

\section{Conclusion}
\ \ We have derived the hermitian time operator for non-relativistic quantum mechanics$(5)$.We have then checked it for consistency with quantum mechanical formalisms in section 3.Application to the free particle system has shown that the particle persists in motion for eternity.We have shown that backward flow of time is impossible at least locally.This means that quantum mechanics does not require backward flow of time in the explanation of delayed choice experiments.We have made a prediction that electron transition between states is instantaneous assuming the hermitian operator for the non-relativistic case.

\section{Acknowledgements}
We acknowledge \textbf{Lexy Andati} for services offered in the typing of this paper.

\newpage
\section{References}
\ \ \ \ \ [1] Lange O.L,Raab R.E,Operator Methods in Quantum Mechanics 

\ \ \ \ (Oxford University Press,New York) 1991
pp20,21,24,35\\ 

[2] Muga J,Mayato R,Egusquiza I, Time in Quantum Mechanics(Springer,Berlin)2008 

\ \ \ \ pp21,109\\

[3] Jewett J.W,Serway R.A, Physics for Scientists and Engineers,7th

\ \ \ \ Ed(Thomson, Belfron California) 2008

\ \ \ \ pp 103,174,1121,1223 \\

[4] Kreyszig E,Advanced Engineering Mathematics, 9th Ed.

\ \ \ \ (John Wiley and Sons Hoboken New Jersey)

\ \ \ \ pp 407,408\\

[5] Brigitte Falkenburg, Correspondence Principle 2009 Co: Springer Link

\ \ \ \ Retrieved from: link.springer.com/chapter/10.1007 On 2nd Jan 2015 \\

[6] Richard Fitzpatrick,Heisenberg Equation of Motion,University of

\ \ \ \ \ Texas. Retrieved from: farside.ph.utexas.edu/teaching/am/lectures/node3t.html

\ \ \ \ on 10th Dec,2014\\ 

[7] Rynasiewicz, Robert, Newton's Views on Space,Time, and Motion ,

\ \ \ \ The Stanford Encyclopedia of Philosophy (Summer 2014 Edition),

\ \ \ \ Edward N.Zalta  (ed.) 

\ \ \ \  ¡http://plato.stanford.edu/archives/sum2014/entries/newton-
stm/¿.\\ 

[8] Jain K.V, Introduction to Quantum Mechanics

\ \ \ \ (Alpha Science International L.T.D. Oxford) 2010

\ \ \ \ pp26,28,48,54-56,90,298,299
\\ 

[9] Chiatti L,\textit{Is Bohr's challenge still relevant?}

\ \ \ \ \ arXiv:1412.3447v1 [physics.gen-ph]

\end{document}